\documentclass[twocolumn, superscriptaddress, floatfix, showpacs, aps, prb,10pt]{revtex4-1}
\pdfoutput=1
\usepackage{graphicx}
\usepackage{amsmath}
\usepackage[colorlinks=true, citecolor=blue, urlcolor=blue]{hyperref}
\usepackage{amssymb}
\usepackage{bm}
\usepackage{color}
\usepackage{array}
\usepackage{booktabs}
\usepackage{multirow}

\DeclareMathOperator{\Tr}{Tr}
\newcommand{\abs}[1]{\lvert#1\rvert}

\begin{document}

\title{Statistical Topological Insulators}
\author{I. C. Fulga}
\affiliation{Instituut-Lorentz, Universiteit Leiden, P.O. Box 9506, 2300 RA Leiden, The Netherlands}
\author{B. van Heck}
\affiliation{Instituut-Lorentz, Universiteit Leiden, P.O. Box 9506, 2300 RA Leiden, The Netherlands}
\author{J. M. Edge}
\affiliation{Instituut-Lorentz, Universiteit Leiden, P.O. Box 9506, 2300 RA Leiden, The Netherlands}
\author{A. R. Akhmerov}
\affiliation{Instituut-Lorentz, Universiteit Leiden, P.O. Box 9506, 2300 RA Leiden, The Netherlands}
\affiliation{Department of Physics, Harvard University, Cambridge, Massachusetts 02138 USA}

\date{October 2013}

\begin{abstract}
We define a class of insulators with gapless surface states protected from localization
due to the statistical properties of a disordered ensemble,
namely due to the ensemble's invariance under a certain symmetry.
We show that these insulators are topological, and are protected by a $\mathbb{Z}_2$ invariant.
Finally, we prove that every topological insulator gives rise to an infinite number of classes of
statistical topological insulators in higher dimensions. Our conclusions are confirmed by numerical simulations.
\end{abstract}
\maketitle

\section{Introduction}

One common definition of a topological insulator (TI) is
that it is a bulk insulator with a gapless surface Hamiltonian which cannot be continuously transformed into a gapped one.\cite{Hasan2010, Qi2011} The surface states of a TI are protected from Anderson
localization and, since there is an anomaly associated
with the TI bulk field theory,\cite{Qi2008, Ryu2012, Wang2011} they
are also robust against interactions as long as the latter
respect the discrete symmetry of the system. Other possible
descriptions of TIs arise from nonlinear sigma-models,\cite{Ryu2010} K-theory,\cite{Kitaev2009} Green's functions,\cite{Volovik2003,Gurarie2011,Essin2011} and even string theory.\cite{Ryu2010a}

There are, however, several known examples of disordered systems whose
surface has a Hamiltonian that can be continuously deformed into a gapped one, and
yet is protected against Anderson localization. One such example is a
so-called weak TI, a 3D material made by
stacking many layers of a 2D TI. Its surface
has two Dirac cones which can be coupled by a mass term, producing a gapped
system. Nevertheless, Ringel~\emph{et al.} have argued in Ref.~\onlinecite{Ringel2012} that since an odd number of weak TI layers is conducting, its surface must always be
metallic. This prediction was tested numerically \cite{Mong2012} and later
explained \cite{Fu2012} in terms of $\mathbb{Z}_2$ vortex
fugacity of a corresponding field theory. Another example is a TI subject to a random magnetic field which is zero on average:\cite{Nomura2008} a random sign gap appears in the surface dispersion, driving the surface to a critical point of the Chalker-Coddington network model.\cite{Chalker1988}

These two examples share one common trait. In order for the surface to avoid
localization, the disordered ensemble must be invariant under a certain
symmetry: translation for a weak TI or
time-reversal for a strong TI with a random magnetic field. We show that this property defines a broad
class of systems, which we call statistical topological insulators (STI).
An STI is an ensemble of disordered systems belonging to the same symmetry class. This ensemble, as a whole, also has to be invariant under an extra symmetry, which we call statistical symmetry since it is not respected by single ensemble elements. These elements have surfaces pinned to the middle of a topological phase transition and protected from localization due to the combined presence of the statistical symmetry and the symmetry of each element, if any. For example, for a weak TI the statistical symmetry is translation, while the symmetry of each element is time-reversal.

Some STIs without disorder become topological crystalline insulators,
introduced by Liang Fu,\cite{Fu2011, Hsieh2012} since they have a gapless
surface dispersion protected by their crystalline symmetry. Nevertheless, not all topological crystalline insulators become STIs once
disorder is added, and the ensemble symmetry need not be crystalline, as in the case of a TI in a random magnetic field.

\begin{table}[t]
\setlength{\tabcolsep}{6pt}
\setlength\extrarowheight{3pt}
\begin{tabular}{ c | l l l l | l l l l }
 \hline \hline
 \multirow{3}{1.5cm}{\centering{Symmetry class}} & \multicolumn{4}{c|}{TI}  & \multicolumn{4}{c}{STI}\\ 
  & \multicolumn{4}{c|}{$d$}  & \multicolumn{4}{c}{$d$}\\ 
  & 1 & 2 & 3 & 4 & 1 & 2 & 3 & 4 \\
  \hline
  A & $\,$- & $\mathbb{Z}$ & $\,$- & $\mathbb{Z}$ & $\,$- & $\,$- & \checkmark & \checkmark \\
  AIII & $\mathbb{Z}$ & $\,$- & $\mathbb{Z}$ & $\,$- & $\,$- & \checkmark & \checkmark & \checkmark \\
  \hline
  BDI & $\mathbb{Z}$ & $\,$- & $\,$- & $\,$- & $\,$- & \checkmark & \checkmark & \checkmark \\
  D & $\mathbb{Z}_2$ & $\mathbb{Z}$ & $\,$- & $\,$- & $\,$- & \checkmark & \checkmark & \checkmark \\
  DIII & $\mathbb{Z}_2$ & $\mathbb{Z}_2$ & $\mathbb{Z}$ & $\,$- & $\,$- & \checkmark & \checkmark & \checkmark \\
  AII & $\,$- & $\mathbb{Z}_2$ & $\mathbb{Z}_2$ & $\mathbb{Z}$ & $\,$- & $\,$- & \checkmark & \checkmark \\
  CII & $\mathbb{Z}$ & $\,$- & $\mathbb{Z}_2$ & $\mathbb{Z}_2$ & $\,$- & \checkmark & \checkmark &\checkmark \\
  C & $\,$- & $\mathbb{Z}$ & $\,$- & $\mathbb{Z}_2$ & $\,$- & $\,$- & \checkmark & \checkmark \\
  CI & $\,$- & $\,$- & $\mathbb{Z}$ & $\,$- & $\,$- & $\,$- & $\,$- & \checkmark \\
  AI & $\,$- & $\,$- & $\,$- & $\mathbb{Z}$ & $\,$- & $\,$- & $\,$- & $\,$- \\
  \hline\hline
\end{tabular}
\caption{Comparison of combinations of dimension $d$ and symmetry class that allow for non-trivial TIs (left) and STIs (right). The left part of the table shows the original classification of TIs\cite{Ryu2010,Kitaev2009}. In the right part of the table, ticks mark combinations of symmetry class and dimensionality which allow for STIs. STIs require that $d\geq 2$ and that there exists a TI in the same symmetry class in $d'$ dimensions, with $d'<d$. For $d>4$ an STI phase is possible in all symmetry classes. \label{STI_periodic_table} }
\end{table}

We show that STIs are a true bulk phase: in order for
the surface to become localized without breaking the symmetries, the bulk must
undergo a phase transition. Since the bulk transition of an STI is a topological phase
transition by itself, it is possible to construct a higher dimensional system
with its surface pinned to the middle of an STI phase transition. Such a
construction makes every single topological phase transition \cite{Kitaev2009, Ryu2010}
give rise to infinitely many higher-dimensional descendant topological
phases, as shown in Table \ref{STI_periodic_table}. The metallic phases of a triangular Majorana lattice,\cite{Kraus2011,Laumann2012} which we discuss below, and of a 2D system with sublattice symmetry,\cite{Gade1991,Gade1993,Motrunich2002} are examples of STIs with two statistical symmetries, in both cases either reflection or translation symmetries.

The manuscript has the following structure. In Section \ref{sec:STI_def} we start by defining the STI topological invariant in the case of a $\mathbb{Z}_2$ statistical symmetry group. In section \ref{sec:STI_reflection_symm} we show how to build a tight-binding model for an STI
in any dimension and symmetry class, using statistical reflection symmetry. Finally, we check the properties of STIs numerically in Sec.~\ref{sec:numerics}. We conclude in Sec.~\ref{sec:conclusions}.

\section{Construction of an STI}
\label{sec:STI_def}

% In this section we first discuss the general requirements for an STI and how an average symmetry can prevent a surface from localization.

To determine the necessary conditions required to obtain an STI, let us consider an ensemble of $d$-dimensional systems with $(d\!-\!1)$-dimensional surface. We require that the Hamiltonian $H_i$ of each ensemble element be local, belong to the same symmetry class, and that the correlation function of the Hamiltonian matrix elements be sufficiently short-ranged. Additionally, we require that the bulk be insulating. The surface should have a combination of dimensionality and symmetry class allowing it to be in a topological phase with invariant $Q_{d-1}$. For example, if the surface is two-dimensional and in  symmetry class A (neither time reversal, chiral, nor particle-hole symmetry are present), $Q_{d-1}$ is the Chern number. We consider $d \geq 2$, so that both surface and bulk are self-averaging.\footnote{Nevertheless, a 1D precursor to the STI phase transition can be found in a half-filled Anderson chain\cite{Kappus1981}} Finally, the ensemble should also possess a statistical symmetry. This means that 
every ensemble element $H_i$ is equally likely to appear as $\mathcal{U}H_i\mathcal{U}^{-1}$, with $\mathcal{U}$ a unitary or anti-unitary operator. Examples of such symmetries are reflection, inversion, and time-reversal. Alternatively $\mathcal{U}$ can represent a statistical anti-symmetry, such as particle-hole or chiral symmetry. In this case $H_i$ appears equally likely as $-\mathcal{U}H_i\mathcal{U}^{-1}$, its e.g. particle-hole reversed partner.

\subsection{Identification of an STI topological invariant for the ensemble}
\label{sec:basic-sti-constr}

\begin{figure}[tb]
\begin{center}
 \includegraphics[width=\columnwidth]{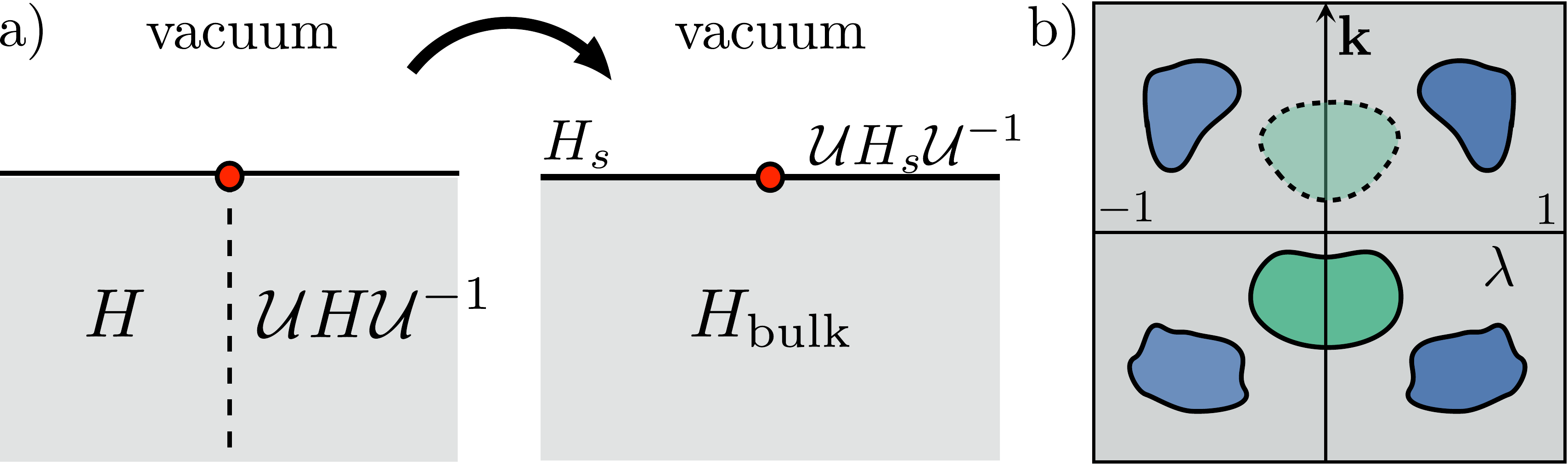}
 \caption{(a): Given two disordered systems $H_i, \mathcal{U}H_i\mathcal{U}^{-1}$ mapped into each other by a symmetry $\mathcal{U}$ and joined together, the presence of gapless states at the common surface interface (red circle) reveals a difference in their surface topological invariant. The presence of gapless states can be determined by deforming the system into a clean one with a common, $\mathcal{U}$-symmetric $H_\textrm{bulk}$ and a domain wall between two $\mathcal{U}$-symmetry-broken surfaces. (b): Examples of possible patterns of surface Fermi surfaces as a function of the parameter $\lambda$, interpolating between a surface and its symmetry reverse, and the surface momenta $\bm{k}$. Green and blue puddles represent Fermi surfaces located at the symmetric point $\lambda=0$ or away from it, respectively. A non-trivial (trivial) STI has an odd (even) number of Fermi surfaces at the symmetric point, not counting Kramers degeneracy. Thus, in this example, the STI invariant $\mathcal{Q} $ takes value $-
1 (+1) $ in the absence (presence) of the dashed Fermi surface.
\label{fig:fermi_surfaces}}
\end{center}
\end{figure}

Let us now show how it is possible to identify a bulk topological invariant for such an ensemble of disordered Hamiltonians. We consider an interface between two ensemble elements, $H_i$ and $\pm\mathcal{U}H_i\mathcal{U}^{-1}$, shown in Fig. \ref{fig:fermi_surfaces}(a). This combined system is also an ensemble element and hence its bulk is insulating, since all elements of the ensemble have insulating bulk.
Furthermore, if the surfaces of $H_i$ and $\pm\mathcal{U}H_i\mathcal{U}^{-1}$ are also insulating, then due to self-averaging they share the same topological properties. Hence the $(d-2)$-dimensional boundary separating the surfaces should carry no topologically protected gapless states. Our aim is to show that for certain ensembles the number of such states must be non-zero, thus contradicting the assumption of an insulating surface.

The $(d-2)$-dimensional boundary can be viewed as a topological defect arising at the interface between the two systems described by $H_i$ and $\pm\mathcal{U}H_i\mathcal{U}^{-1}$, shown as a red dot in Fig.~\ref{fig:fermi_surfaces}(a). Provided these Hamiltonians vary slowly away from the defect, the number of topologically protected gapless states occurring at the boundary can be computed by considering an adiabatic path surrounding it, as described by Teo and Kane in Ref.~\onlinecite{Teo2010}. Instead, we deform our system into a simpler one [see Fig. \ref{fig:fermi_surfaces}(a)]. We first add a translationally-invariant surface term $H_s\equiv \mp\mathcal{U} H_s \mathcal{U}^{-1}$ to $H_i$ which strongly breaks the statistical symmetry and gaps the surface. For instance, if the statistical symmetry is time-reversal, $H_s$ could be a strong, uniform Zeeman field at the surface. Simultaneously, a term $-H_s\equiv\pm\mathcal{U} H_s \mathcal{U}^{-1}$ is added to $\pm\mathcal{U}H_i\mathcal{U}^{-1}$, on the 
other side of the interface. If in the process of adding $H_s$ to $H_i$ the surface gap closes, so does the surface gap on the symmetry-reversed side.
Thus, the parity of the number of topologically protected gapless states at the boundary does not change. Then, we deform $H_i$ and $\pm\mathcal{U}H_i\mathcal{U}^{-1}$ to remove
both disorder and symmetry-breaking, taking care that in the process the gap does not close anywhere: neither in the bulk of $H_i$ and $\pm\mathcal{U} H_i \mathcal{U}^{-1}$ nor at their interface. The new bulk Hamiltonian $H_\textrm{bulk}$ has no disorder and satisfies $H_\textrm{bulk}=\pm\mathcal{U} H_\textrm{bulk}\mathcal{U}^{-1}$. For weak disorder, this step corresponds to reducing disorder strength to zero. While finding $H_\textrm{bulk}$ in the case of strong disorder is non-trivial, we do not know of any obstructions which would make it impossible, nor of any counter-examples.

The evaluation of the number of topologically protected boundary states is straightforward in this new system, since it amounts to studying a domain wall between two clean, gapped surfaces. This number is given by the difference in surface topological invariant, $\Delta Q_{d-1}$, which can be computed by standard methods, see e.g.~Ref.~\onlinecite{Essin2011}, applied to the Hamiltonian $H= H_\textrm{bulk}+\lambda H_s$, with $\lambda\,\in\,[-1,1]$. In Fig. \ref{fig:fermi_surfaces}(b) we show two possible configurations of gap closings of $H$ in $(\bm{k},\lambda)$ space, with $\bm{k}$ the momentum parallel to the surface. Since a gap closing at a finite value of $\lambda$ always has a partner at $-\lambda$, the parity of $\Delta Q_{d-1}$ is dictated by the properties of $H(\bm{k},\lambda=0)\equiv H_\textrm{bulk}$.

If $\Delta Q_{d-1}$ is non-zero, $H_i$ and $\pm\mathcal{U}H_i\mathcal{U}^{-1}$ must have different surface topological invariants. At the same time, however, the combined system of $H_i$ and $\pm\mathcal{U}H_i\mathcal{U}^{-1}$ should have the same surface topological invariant as $H_i$ and $\pm\mathcal{U}H_i\mathcal{U}^{-1}$ individually, for sufficiently large system sizes. This is dictated by the fact that topological invariants are self-averaging in insulating phases, since they are measurable quantities related to response coefficients. We then see that, since $H_i$ and $\pm\mathcal{U}H_i\mathcal{U}^{-1}$ are equally probable ensemble elements, a contradiction arises.

In particular, if $\Delta Q_{d-1}$ is odd, the contradiction cannot be avoided, because the number of topologically protected gapless states must be different from zero. The only possible way out is to conclude that in this case the surface cannot be insulating. On the other hand, if $\Delta Q_{d-1}$ is even, the ensemble symmetry does not prevent the insulating phase from appearing.

We thus define $\mathcal{Q} = (-1)^{\Delta Q_{d-1}}$ as the $\mathbb{Z}_2$ topological invariant of an STI. The STI topological invariant $\mathcal{Q}$ is a bulk property, e.g. the parity of the mirror Chern number. Nevertheless, the evaluation of $\mathcal{Q}$ for large disorder strength is in general a hard problem, since it is necessary to find a symmetric $H_\textrm{bulk}$ which can be connected to an $H_i$ without closing the bulk gap.

\subsection{Higher dimensional generalizations}
\label{sec:high-dimens-gener}

This construction can be repeated recursively, by considering an ensemble of $(d+1)$-dimensional systems with a $d$-dimensional surface and a second statistical symmetry $\mathcal{U}_2$ in addition to $\mathcal{U}_1$. The surfaces of the ensemble elements, if gapped, possess a $d$-dimensional STI invariant $\mathcal{Q}$, protected by the statistical symmetry $\mathcal{U}_1$. We may now ask whether protected gapless states appear at a $(d-1)$-dimensional boundary between the surfaces of two ensemble elements $H^{(d+1)}_i$ and $\pm\mathcal{U}_2H^{(d+1)}_i\mathcal{U}_2^{-1}$. In other words, we want to know if the boundary between the surfaces of $H^{(d+1)}_i$ and $\pm\mathcal{U}_2H^{(d+1)}_i\mathcal{U}_2^{-1}$ is itself a protected surface of an STI. The problem can then again be reduced to the study of the gap closings of a clean Hamiltonian $H^{(d+1)}=H^{(d+1)}_\textrm{
bulk}+\lambda_2H_{s_2}$, where $\lambda_2\,\in\,[-1,1]$ and $H_{s_2}$ strongly breaks the statistical symmetry $\mathcal{U}_2$ but commutes with $\mathcal{U}_1$. The parity of the change $\Delta \mathcal{Q}$ of the $d$-dimensional STI invariant is determined by the gap closings at $\lambda_2=0$. If $\Delta\mathcal{Q}$ is odd, then topologically protected states must appear at the interface between the two surfaces, contradicting the assumption that the surfaces are gapped and topologically equivalent. Hence, the ensemble must have gapless surfaces, protected from localization by the combined presence of $\mathcal{U}_1$ and $\mathcal{U}_2$. By repeatedly adding more symmetries and dimensions it is possible to construct STIs in dimension $d+n$ using an ensemble $\mathbb{Z}_2^n$ symmetry and a $d$-dimensional topological invariant.

\section{STI models with reflection symmetry}\label{sec:STI_reflection_symm}

To illustrate the general idea presented in the previous Section, we now show that ensemble reflection symmetry allows us to construct a $d$-dimensional STI in any symmetry class which allows a non-trivial invariant in $(d-1)$ dimensions.

\begin{figure}
\includegraphics[width=\linewidth]{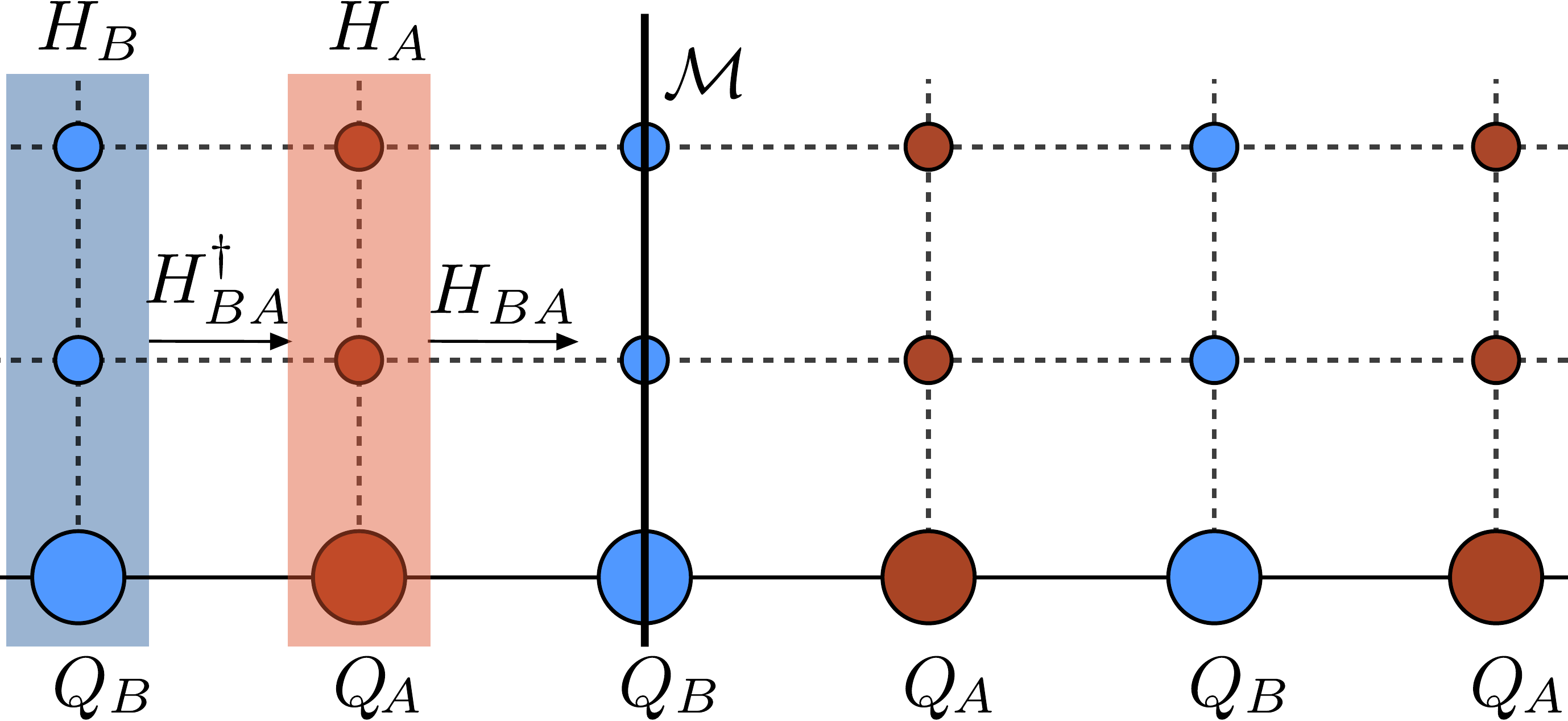}
\caption{A system consisting of infinitely many layers of two different types, $A$ (red) and $B$ (blue), with  Hamiltonians $H_A$ and $H_B$ and with staggered topological invariants $Q_A$ and $Q_B$. Sites within the layer are marked by circles, with bigger ones denoting the end of each layer. Hopping from the $A$-layers to $B$-layers equals $H_{BA}$. In the absence of disorder, the system is translationally invariant, with a unit cell composed of two
layers. It also possesses a reflection symmetry $\mathcal{R}$ with respect to an axis passing through one layer (black line).\label{fig:stack}}
\end{figure}

Let us consider a $d$-dimensional system consisting of an infinite stack of $(d-1)$-dimensional weakly coupled layers (see Fig.~\ref{fig:stack}). In particular, we consider alternating layers of two types, $A$ and $B$, with Hamiltonians $H_A$ and $H_B$. The hopping from layer $A$ to layer $B$ along both the positive and negative stacking direction equals to $H_{BA}$. We require that $H_A$ and $H_B$ both be gapped in the bulk and $H_{BA}$ be smaller than the bulk gap of each of the layers. Under these conditions the bulk of the complete stack also stays gapped.

% The layers have a staggered topological invariant.
We consider a geometry where each layer is semi-infinite along one spatial dimension and infinite along the remaining ones. The edge of each layer has $(d-2)$-spatial dimensions, so that the whole stack has a $(d-1)$-dimensional surface. The system belongs to a symmetry class which allows the layers to have a non-trivial topological invariant with values $Q_A$ and $Q_B$.
We require $Q_B$ to be the inverse group element of $Q_A$ (i.e. $Q_B=-Q_A$ for $\mathbb{Z}$ invariants and $Q_B=Q_A$ for $\mathbb{Z}_2$), so that layers of $A$ and $B$ type have an equal number of topologically protected gapless states at the surface. Due to this condition a pairwise coupling of the layers makes the gapless states at the surface gap out, resulting in a topologically trivial system. For instance, in the unitary symmetry class with $d=2$, $H_B$ may be the time-reversed partner of $H_A$, with an opposite Chern number. The alternating layers then support $\abs{Q_A}$ chiral edge states propagating in opposite directions.

% The system has a reflection symmetry, which becomes ensemble symmetry with disorder.
By construction, such a model has a reflection symmetry with respect to an axis passing through any of $A$-layers, with operator
\begin{equation}
  \label{eq:refl_sym}
  \mathcal{R} =
  \begin{pmatrix}
    1 & 0 \\
    0 & e^{-ik}
  \end{pmatrix},
\end{equation}
acting on the Bloch wave function
\begin{equation}
\psi_k(x) = \exp(ikx)\,\begin{pmatrix} \psi_A \\  \psi_B\end{pmatrix}\,.
\end{equation}
It is straightforward to verify that the Bloch Hamiltonian
\begin{equation}
  \label{eq:bloch_ham}
  H(k) =
  \begin{pmatrix}
    H_A & H_{BA}^\dagger (1 + e^{-ik}) \\
    H_{BA}(1 + e^{ik}) & H_B
  \end{pmatrix}
\end{equation}
is reflection-symmetric, i.e. that it obeys
\begin{equation}
\mathcal{R}\,H(k)\,\mathcal{R}^\dagger=H(-k)\,.
\end{equation}
Adding any local disorder to $H_A$ and $H_B$ makes $\mathcal{R}$ an ensemble symmetry. 

% To understand statistical topological properties, count the FS in the clean limit.
In the previous Section we have argued that for sufficiently weak disorder the behavior of a disordered system with a statistical symmetry is dictated by the parity of the number of gap closings at the surface of the system in the absence of disorder. Therefore, for our purposes it will be sufficient to to determine the number of surface gap closings in the clean system that are protected by the reflection symmetry. Since the Bloch Hamiltonian \eqref{eq:bloch_ham} does not couple $\psi_A$ with $\psi_B$ at $k=\pi$, the number of zero energy eigenstates of $H(\pi)$ is equal to the combined number of topologically protected states of $H_A$ and $H_B$, i.e. it is equal to $2\abs{Q_A}$. Therefore, $H(k)$ will possess $\abs{Q_A}$ (non-chiral) Fermi surfaces centered around $k=\pi$.

 For weak disorder, the STI invariant can be computed by choosing Eq.~\eqref{eq:bloch_ham} as $H_\textrm{bulk}$ and adding a reflection symmetry-breaking term $\lambda H_s$. An example of $H_s$ is the term that doubles every even hopping in a large region near the surface, and removes every odd hopping. The Bloch form of such a term is
\begin{equation}
  H_s=
  \begin{pmatrix}
    0 & H_{BA}^\dagger(1 - e^{-ik}) \\
    H_{BA}(1 - e^{ik}) & 0
\end{pmatrix}.
\end{equation}
Since this term fully gaps the surface, and since there are $\abs{Q_A}$ non-chiral Fermi surfaces at the symmetric point, we conclude that for $\abs{Q_A}$ odd, a disordered stack of such layers is an STI. Hence, if each layer originally carried an odd number of topologically protected edge states, the layered system is a nontrivial STI. This procedure allows one to construct tight-binding models showing an STI phase in any symmetry class.

\section{Numerical Simulations}\label{sec:numerics}

We have used the Kwant code \cite{Groth2013} to perform numerical checks of our predictions.\footnote{The source code of the numerical calculations is available online at \dots}
As an example of a 2D STI we consider a stack of coupled Kitaev
chains \cite{Kitaev2001} (symmetry class D). The 2D lattice Hamiltonian of this system has the form
\begin{equation}
H_\textrm{D} = (2t_y \cos k_y - V ) \tau_z + \Delta \tau_y \sin k_y + \alpha \tau_x \sin k_x, \label{eq:classD_H}
\end{equation}
where $\tau_i$ are Pauli matrices in Nambu space, $x,y$ are integer coordinates perpendicular and along the chain direction, $t_y=1$ is the normal hopping, $V$ the onsite disorder potential uniformly distributed in the interval $[-\delta/2, \delta/2]$, $\Delta=1$ the $p$-wave pairing strength within each chain, and $\alpha=0.45$ the interlayer coupling strength. This Hamiltonian has particle-hole symmetry, $H_\textrm{D}=-\tau_x H_\textrm{D}^\ast \tau_x$, and is related to a Hamiltonian with an ensemble reflection symmetry by a gauge transformation $\psi(x) \rightarrow (-1)^{\lfloor x / 2 \rfloor} \psi(x)$.

\begin{figure}[tb]
\includegraphics[width=\linewidth]{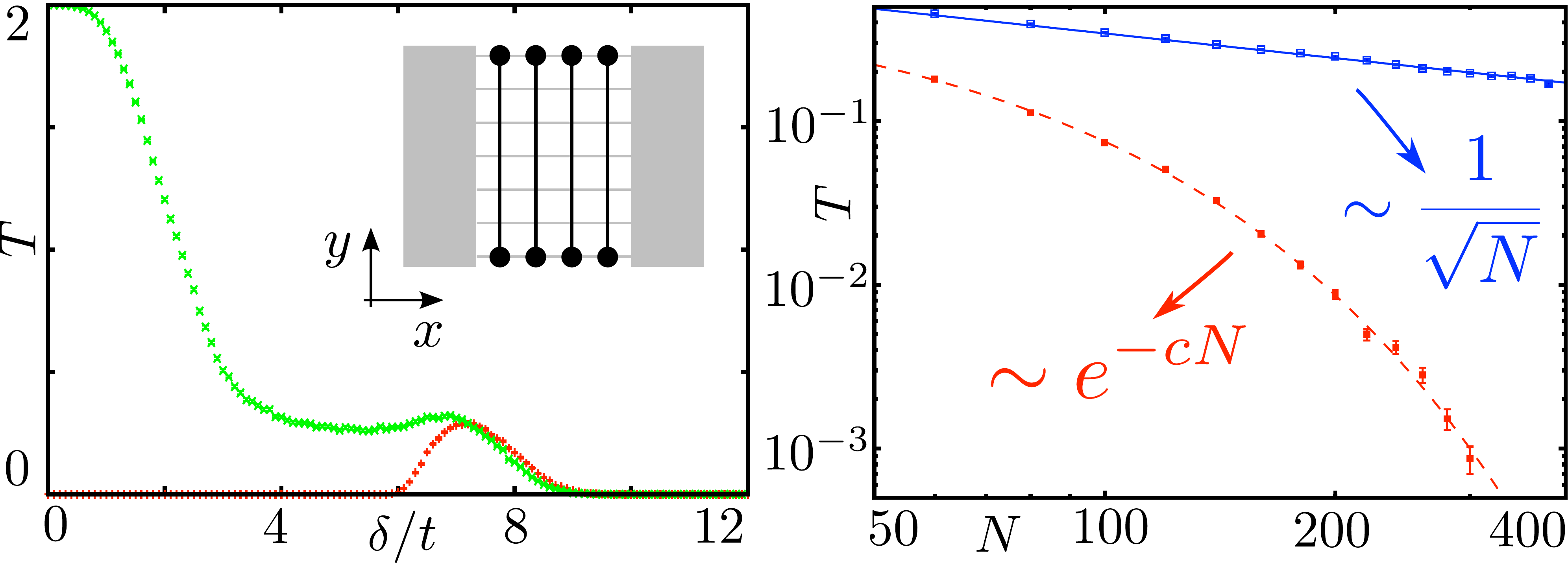}
\caption{\emph{Left:} Transmission $T$ versus disorder strength $\delta$ through a stack of 49 coupled Kitaev chains, each 150 sites long, with hard wall (green) and periodic (red) boundary conditions, averaged over approximately $6 \times 10^3$ disorder realizations. Error bars are smaller than symbol sizes. The appearance of a finite transmission at high disorder strength in the case of periodic boundary conditions is a consequence of the bulk gap closing, which accompanies an STI-trivial insulator transition. The inset shows the sample geometry: Kitaev chains are black lines, leads are gray rectangles. \emph{Right:} Transmission as a function of number of Kitaev chains (blue, solid line) or of number of BDI chains with $Q=\pm 2$ (red, dashed line) in a stack, averaged over approximately $2\times 10^4$ disorder realizations with $\delta = 3t$ and chain length of $40$ lattice sites. \label{fig:kitaev_gq}}
\end{figure}

We have attached ideal leads to a stack of Kitaev chains along the $x$-direction, using periodic or hard-wall boundary conditions in the $y$-direction.
In Fig.~\ref{fig:kitaev_gq} we show the calculated total quasiparticle transmission between the leads, $T=\Tr (t^\dagger t)$ with $t$ the transmission block of the scattering matrix.
The clean system with $\delta=0$ and hard wall boundary conditions has transmission $T=2$ due to a non-chiral Majorana mode at each edge. 
This transmission is reduced by disorder, however it only disappears after the bulk goes through a delocalization transition to become a trivial Anderson insulator, as shown by the transmission with periodic boundary conditions.

To test the properties of the transmitting edges, we calculate the dependence of $T$ on the number of Kitaev chains $N$ at a fixed $\delta$. To verify that it is the parity of the number of gap closings that determines whether an ensemble is an STI or not, we compare this behavior to that of a stack of chains in symmetry class BDI with alternating topological invariants $Q=\pm 2$ (see Fig.~\ref{fig:kitaev_gq}). The two-dimensional BDI Hamiltonian reads:

\begin{eqnarray}\label{eq:classBDI_H}
 H_{\textrm{BDI}} &=& (2t_y \cos k_y - V) \sigma_0\tau_z + A \sigma_z\tau_z + B\sigma_x\tau_z \nonumber\\
         & & + C \sigma_y\tau_y - \Delta \sigma_0 \tau_y \sin k_y - \alpha \sigma_z\tau_y \sin k_x,
\end{eqnarray}
with $\sigma_i$ and $\tau_i$ Pauli matrices acting on the time-reversal and particle-hole degrees of freedom, respectively. The Hamiltonian \eqref{eq:classBDI_H} obeys particle-hole symmetry, $H_\textrm{BDI}=-\tau_xH_\textrm{BDI}^\ast\tau_x$, as well as time-reversal symmetry $H_\textrm{BDI}=H_\textrm{BDI}^\ast$. The alternating topological invariants are obtained by staggering the sign of the $p$-wave pairing strength on each chain, $\Delta \rightarrow  (-1)^{x}\,\Delta$, and the total quasiparticle transmission is obtained for $A=0.4$, $B=C=0.3$, and all other parameters the same as for the class D Hamiltonian \eqref{eq:classD_H}.

We find that $T \sim N^{-1/2}$ for the stack of Kitaev chains, as expected for a chain of randomly coupled Majorana bound states (MBS), or more generally for 1D systems at the critical point \cite{Brouwer2000,Brouwer2003,Motrunich2001,Gruzberg2005}, while the edges of the BDI stack are localized with $T \sim \exp(-cN)$.

To test STIs in a dimension two higher than the dimension of the original topological invariant, we consider a triangular lattice of MBS,\cite{Biswas2011,Kraus2011,Laumann2012} which is a surface model of a 3D array of coupled Kitaev chains. The tight-binding Hamiltonian of this model is given by
\begin{align}
  H_\bigtriangleup=\sum_{\langle ij \rangle} i t_{ij}\gamma_i\gamma_j\,,
  \label{eq:1}
\end{align}
with real $t_{ij}=-t_{ji}$. If in a clean translationally invariant system with one MBS per unit cell hoppings have equal magnitude, then the system has a reflection symmetry with respect to a plane passing through one of the hoppings and a reflection anti-symmetry with respect to a perpendicular plane passing through any site. There is one Fermi surface in the clean system, hence any disorder that preserves the two reflection symmetries on average should make this lattice of MBS a surface model of an STI, with STI invariant equal to the parity of the number of MBS per unit cell.

\begin{figure}[tb]
\begin{center}
 \includegraphics[width=\linewidth]{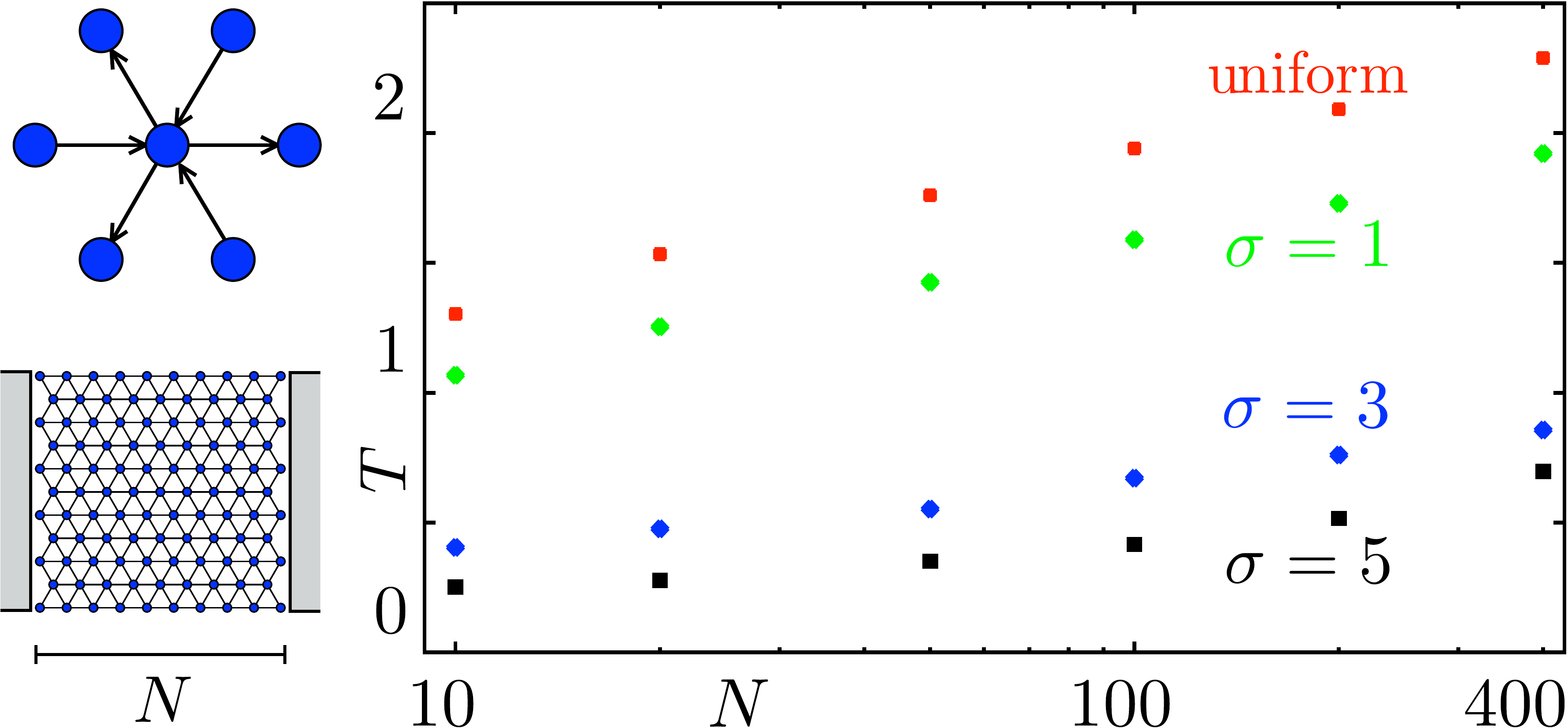}
 \caption{\label{fig:recursiveSTI} \emph{Top left}: Arrows show the directions in a triangular Majorana lattice where average hopping values are chosen to be positive. \emph{Bottom left}: Square-shaped disordered Majorana lattice
 with leads shown as gray rectangles. \emph{Right}: Transmission $T$ between leads versus system size $N$. The red curve corresponds to the case of hoppings uniformly distributed in a range $[-t,2t]$. The other three curves correspond to log-normal distributed hoppings, with zero log-mean and different values of the log-variance $\sigma$.}
\end{center}
\end{figure}

In Fig.~\ref{fig:recursiveSTI} we show that the calculated transmission through a square-shaped region of a disordered Majorana lattice increases with system size for different disorder types and strengths. Our results explain the thermal metal reported in Refs.~\onlinecite{Kraus2011,Laumann2012} for a triangular Majorana lattice with random uncorrelated hopping signs. Since in that case the ensemble has the reflection symmetries described in the previous paragraph, the metallic phase is a consequence of it being a surface model of an STI. 

\begin{figure}[t!]
\includegraphics[width=\linewidth]{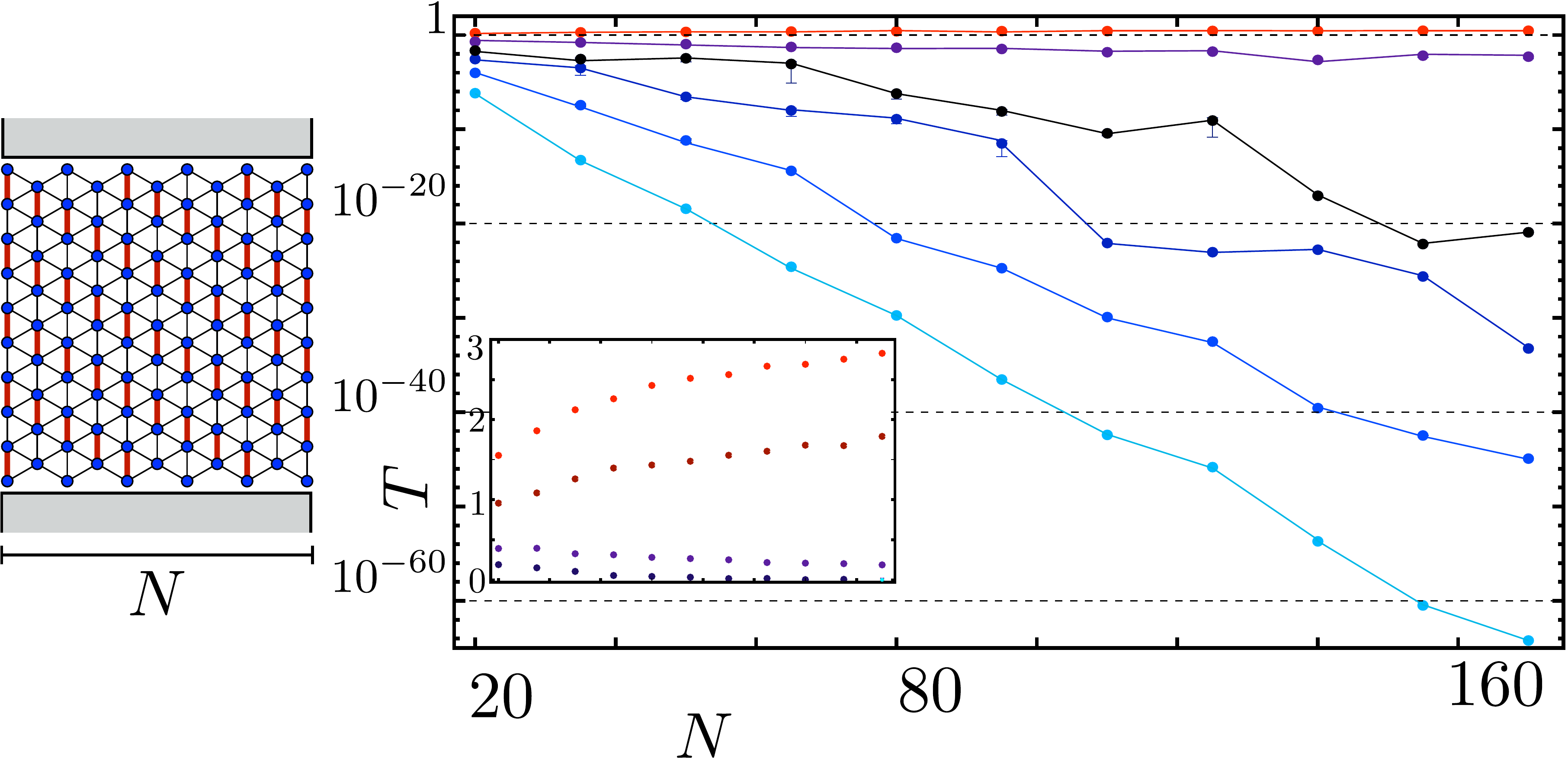}
\caption{\emph{Left}: triangular Majorana lattice with staggered hoppings in the vertical direction. The strength of thin black hoppings are uniformly distributed in the interval $[0.2t, 1.8t]$, while the strengths of thick red hoppings are uniformly distributed in the interval $[t_1+0.2t, t_1+1.8t]$ with $t_1>t$, so that they are larger than $t$ on average. Directions in which hoppings are chosen to be positive are the same as those shown in Fig.~\ref{fig:recursiveSTI}. \emph{Right}: Transmission $T$ as a function of system size $N$, for systems with unit aspect ratio. Different curves correspond to different staggering strengths $t_1/t=2.6, 3.1, 3.3, 3.4, 3.6, 4.2$ with larger values corresponding to lower transmission. The transition from metallic to insulating scaling is also shown in the inset using a linear scale for transmission. There we show the curves $t_1/t=2.6, 2.8, 3.0, 3.1$, with lower transmissions once again corresponding to higher $t_1/t$.}\label{fig:localization}
\end{figure}

As a further confirmation of this topological origin, we have analyzed the same system with broken statistical reflection symmetries. In particular, we have introduced staggered hoppings in the way shown in the left panel of Fig. \ref{fig:localization}, always in the presence of uniform disorder. The staggering of the hoppings breaks the ensemble symmetries that are present in the non-staggered lattice. In agreement with our expectations based on the STI origin of the thermal metal phase, we have observed a transition from metallic to insulating behavior as a consequence of the breaking of the statistical reflection symmetries, as shown in the right panel of Fig. \ref{fig:localization}.

\section{Conclusions and discussion}
\label{sec:conclusions}

We have unified several known examples of systems protected from localization by ensemble symmetries into the new framework of statistical topological insulators. We presented a proof of why STIs avoid localization, by showing that the ensemble symmetry prevents them from having a definite value of a surface topological invariant. We have introduced a universal construction of STIs using reflection symmetry, and were able to explain the thermal metal phase of Refs.~\onlinecite{Kraus2011,Laumann2012} as being a surface model of an STI. Since the identification of the bulk STI invariant with a gap closing relies on the regular TI invariant, STIs should be protected from interactions, as long as the interactions do not introduce spontaneous symmetry breaking with long-range correlations, as reported 
in Ref.~\onlinecite{Liu2012} for weak TIs.

A natural extension of our approach would include providing a more complete relation between ensemble symmetry groups and STIs. While we have focused on $\mathbb{Z}_2^n$ symmetries for simplicity, translational symmetries or anti-symmetries must also be sufficient to construct STI, as is the case for the weak TIs. STIs using fractionalized phases may provide a new way to construct fractional TIs. The way the presence of several TI and STI phases in the same symmetry class enriches the phase diagram of Anderson insulators provides another interesting direction to study. Finally, we are as yet unable to solve the problem of efficiently evaluating the STI topological invariant on a general basis. It is sufficiently simple for several classes, e.g. weak TI, but becomes hard for more complicated ensemble symmetries.

\acknowledgments

We would like to thank C. W. J. Beenakker, L. Fu, T. Hyart and C. R. Laumann for useful discussions. This project was supported by the Dutch Science Foundation NWO/FOM, by an ERC Advanced Investigator Grant, and by the EU network NanoCTM. AA was partially
supported by a Lawrence Golub Fellowship.

\bibliography{sti}

\end{document}